# Polarization control proposal for Shanghai deep ultraviolet free electron laser

Tong Zhang[a,b], Haixiao Deng[a,1], Jianhui Chen[a], Zhimin Dai[a], Chao Feng[a,b], Lie Feng[a], Taihe Lan[a], Bo Liu[a], Dong Wang[a], Xingtao Wang[a], Meng Zhang[a], Miao Zhang[a]

[a] *Shanghai Institute of Applied Physics, Chinese Academy of Sciences, Shanghai, 201800, China*
[b] *Graduate University of the Chinese Academy of Sciences, Beijing, 100049, China*

**Abstract:**

In this paper, a fully coherent radiation option with controllable polarization is proposed for Shanghai deep ultraviolet free electron laser (FEL) test facility. Intensive start-to-end simulation suggests that, the two crossed planar undulators which generate the horizontal and vertical linear polarized FEL respectively, should be placed as close as possible for avoiding the polarization performance degradation of the final combined FEL radiation. With the existence of the phase-shifter between the two crossed radiators, Fourier-Transform-Limited output radiation with 100 nJ order pulse energy, 5 ps full pulse length and circular polarization degree above 90% could be achieved.

**1. Introduction**

With the great success of the Linear Coherent Light Source (LCLS), the world's first x-ray free electron laser (FEL) [1], scientists now have the ability to investigate the much smaller and faster regime, such as femtosecond x-ray single-shot diffraction [2] and time-resolved pump-probe experiments [3], etc. On the other hand, the polarization controls of short-wavelength FEL are of great importance and interest, especially on developing novel powerful spectroscopy for probing valence charge, spins and bonding dynamics [4], etc. APPLE-type undulators [5], which have been widely utilized on the third generation synchrotron radiation sources [6], can provide controllable polarization, but it is hard to quickly change the polarity due to the slow mechanical motion of magnets and, thus, to achieve fast helicity switching. However, there is an ever increasing demand by the users for circularly polarized FEL pulses with helicity switching frequency up to kilohertz, especially in the soft x-ray spectral regime.

---

[1] Email: denghaixiao@sinap.ac.cn

One possible solution for fast helicity switching is the technique of crossed planar undulators [7]. The key point of crossed undulator is making the electromagnetic radiation along the horizontal and vertical axis as identical as possible, including the amplitude and the phase. Recently, various approaches using crossed undulators for polarization control have been reported for the LINAC-based high-gain single-pass FELs [8-13] and storage ring-based FEL oscillators [14, 15]. For self-amplified spontaneous emission (SASE) FEL, the best time to place the crossed undulators is just before its power saturation, and then the density modulation induced by the SASE process will nearly keep the same while passing through the crossed undulators. However, the intrinsic spiky structure and the shot-to-shot fluctuation of SASE will cause FEL polarization performance degradation. For the seeded FELs, the density modulation is due to the interaction of the electrons and the external seed laser, and results in stable and fully coherent output radiations. Thus with the crossed undulators technique, excellent polarization control performances are expected in the seeded FELs.

Seeded FELs are attractive ways for fully coherent soft x-ray radiations, and thus more and more seeded FEL projects are proposed in this regime [16-19]. Among them, Shanghai soft x-ray FEL (SXFEL) [20, 21] is a two-stage high-gain harmonic generation (HGHG) [22, 23] with the frequency up-conversion from 270nm to 9nm. Utilizing crossed undulators with resonance at the harmonics of 9 nm radiation as 4.5nm or 3nm afterburner [24], SXFEL holds promising prospects for generating fully coherent radiations with variable polarization in the "water-window" region.

In this paper, the designs of polarization control experiment at Shanghai deep ultraviolet FEL (SDUV-FEL) [25], a test bed for SXFEL, are given, including the start-to-end simulations and the experiment preparations. It shows that 90% circular polarization degree could be achieved at SDUV-FEL experiments, which will serve as a polarization control prototype for the future SXFEL.

2. Polarization control for SDUV-FEL

SDUV-FEL is a 160 MeV test facility. Up to now, SASE [26], HGHG [27] and echo-enabled harmonic generation (EEHG) [28] have been experimentally demonstrated at SDUV-FEL. By using the existing double-modulator structure and adding some minor revisions, it is also suitable for testing the FEL polarization control experiments. The schematic of the system layout is shown on Figure 1.

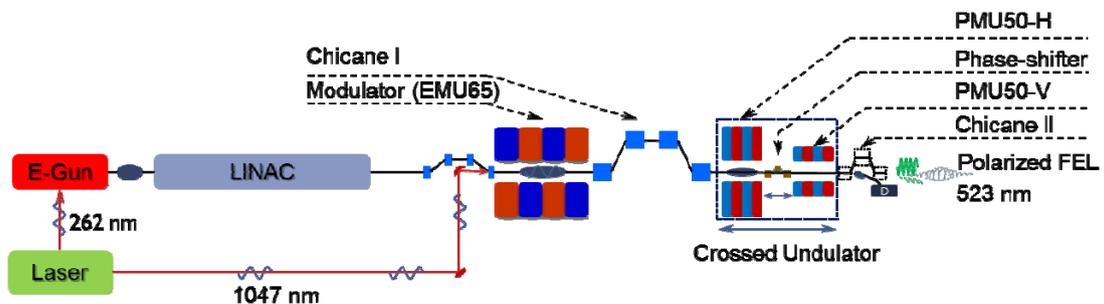

Fig.1. Schematic of SDUV-FEL polarization control mode. The crossed undulators include a pair of permanent planar undulator PMU50-H and PMU50-V, where H and V indicates the horizontal and vertical linear polarized radiation, respectively. A small chicane, so-called "phase-shifter" is used to shift the path delay between the electron and the light.

The polarization control option for SDUV-FEL is proposed to be based on a seeded approach. Firstly, the electron beam is accelerated to 136 MeV, and passes through the first modulator EMU65 and the first dispersive section Chicane I of the EEHG structure, where the density modulation of the electron beam is optimized by interaction with the 10 MW peak power, 1047 nm seed laser. Then the micro-bunched electron beam will emit coherent 523 nm radiation with horizontal linear polarization, i.e. the second harmonic of 1047 nm seed laser, in the second modulator PMU50-H of the EEHG structure. Before entering the long radiator of EEHG, a customized undulator PMU50-V is followed to generate another 523 nm radiation with vertical linear polarization. The density modulation of the electron beam maintains when passing through the two short crossed planar undulators, thus generates two linear polarized radiations with nearly equivalent power and phase, which is important for enhancing the quality of polarized FEL. A small phase-shifter is used for fine tuning of the path difference between the horizontal and vertical linear polarized light, hence to switch the helicity of the combined circularly polarized FEL light.

Table 1 lists the nominal parameters of polarization control mode at SDUV-FEL, which are mainly from the start-to-end simulations and the measurements during the early commissioning stage.

Table 1. Main parameters of polarization control option at SDUV-FEL

| | | |
|---|---|---|
| *Electron Beam* | | |
| Beam Energy | 136 | MeV |
| Slice Energy Spread | 1~2 | keV |
| Normalized Emittance | 4 | mm-mrad |
| Total Charge | 100~250 | pC |
| Peak Current | 50~100 | A |
| *Modulator* | | |
| Resonant Wavelength | 1047 | nm |
| Period Length | 65 | mm |
| Period Number | 10 | |
| *Crossed Undulators* | | |
| Resonant Wavelength | 523 | nm |
| Period Length | 50 | mm |
| Period Number | 10 | |
| *Seed Laser System* | | |
| Wavelength | 1047 | nm |
| Pulse Duration (FWHM) | 8.0 | ps |
| Peak Power | 10 | MW |
| *Phase-shifter* | | |
| $R_{56}$ | 0~2 | μm |
| Phase Shift Range | 0~4π | rad |

3. **Numerical Simulations**

The radiation polarization can be characterized by the two quantities, i.e. the total polarization degree ($P_{tot}$) and circular polarization degree ($P_{cir}$), repectively, define as,

$$P_{tot} = \frac{\sqrt{s_1^2 + s_2^2 + s_3^2}}{s_0}, \quad (1)$$

and

$$P_{cir} = \frac{|s_3|}{s_0}, \quad (2)$$

where $s_0$, $s_1$, $s_2$ and $s_3$ are Stokes parameters that describe the polarization state of electromagnetic radiation [29].

The LINAC beam dynamic results are imported to FEL code GENESIS [30] for the setup shown in Figure 1, i.e., start-to-end simulation, and the FEL output results are illustrated in Figure 2. It demonstrates two Fourier-Transform-Limited output radiation pulses with horizontal and vertical linear polarization, respectively. The full pulse length is about 5 ps and the pulse energy is 100 nJ order. For the combined final radiation, the total polarization degree is rather stable with a level above 90%. It is worth stressing that one can slightly tuning the gaps of crossed undulators to obtain almost equivalent output FEL power, and thus further enhancing the polarization degree of final radiation. However, due to the imperfection of the two crossed linear polarized FEL pulses and the slippage effect between the electron beam and FEL light, the final total polarization cannot reach 100%.

With the phase-shifter, the relative phase difference of the two 523 nm radiations from the crossed undulators can be adjusted, and hence achieving easy helicity switching. Moreover, one can figure out that, 14% peak to peak stability of the electromagnetic phase-shifter is tolerable for a requirement of circular polarization degree larger than 85%, which is fairly loose.

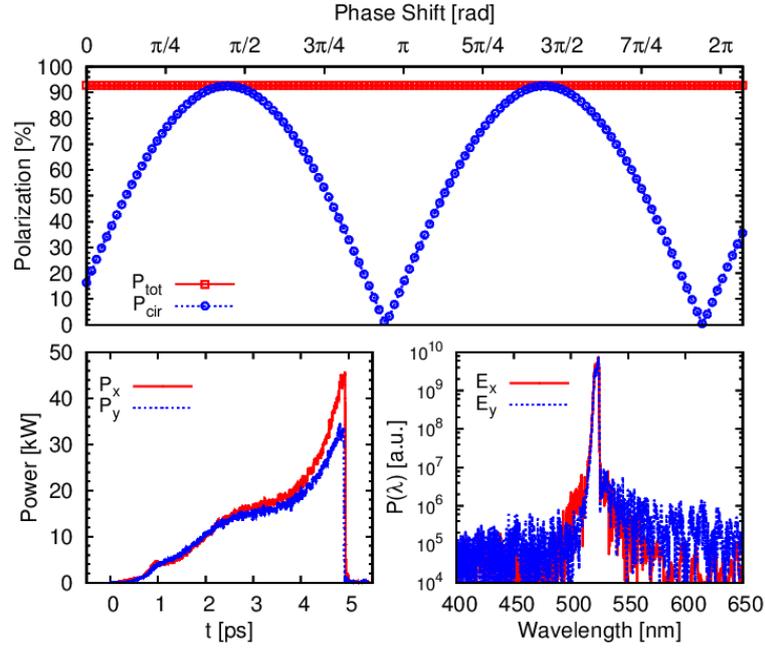

Fig. 2. Start-to-end simulation of FEL polarization control at SDUV-FEL, the upper shows the polarization degree variation as the phase-shifter tuning; the lower illustrates the FEL pulse in the time domain and the frequency domain, respectively.

Generally speaking, to achieve perfect circular polarization, the radiations from the two crossed undulators should be identical and they should transversely and longitudinally overlap with each other. Previous experiments [31, 32] show a local energy spread of 0.5~2 keV for the electron beam of SDUV-FEL, which indicates that the bunching factor of 523 nm radiation could be pretty large. Since the dispersion of the phase-shifter is about 2 microns and the crossed radiators are only 10 periods, the electron beam can be treated as a rigid micro-bunched beam. Therefore, the planar undulator PMU50-H and PMU50-V should supply identical radiations, and we mainly consider the overlap issue, especially the slippage effects in the following discussions.

Because of the electron beam slippage effects before entering the second planar undulator, the longitudinal overlap of FEL radiations emitted by the two crossed undulators will naturally imperfect, which will degrade the polarization performance of the final combined radiation. Figure 3 shows the ELEGANT [33] results of the electron beam longitudinal distribution at the exit of LINAC. Since only S-band accelerating tubes are used in SDUV-FEL, the residual energy chirp induced by radiofrequency curvature is inevitable. The beam energy chirp imprints a phase chirp in the emitted FEL radiations, which will further aggravate the polarization performance together with the slippage effects.

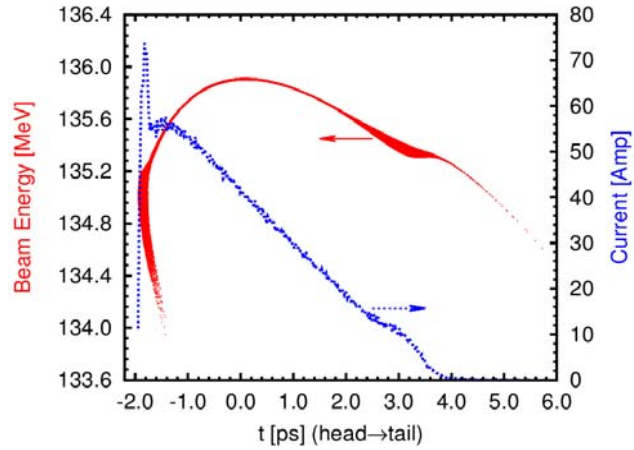

Fig. 3. The longitudinal profiles of the electron bunch after LINAC. The solid and dashed indicate beam energy and beam current, respectively.

In the current configuration of SDUV-FEL, the planar undulator PMU50-H is followed by a 2 m long Chicane II which serves as the second dispersion in EEHG operation mode. To clearly illustrate the beam energy chirp effects on the final FEL polarization performance, intensive start-to-end simulations have been performed for comparison study between two cases, i.e. keeping Chicane II still and placing PMU50-V after Chicane II (Case I), and moving Chicane II backward and placing PMU50-V before Chicane II (Case II). Actually, Case II is the optimized configuration as shown in Figure 1.

Figure 4 gives the radiation phases of the two cases. It is clear that the accumulated polarization degradation induced by the relative phase difference is much smaller in Case II than in Case I. The calculations show that the maximum circular polarization degree is only 60% in Case I, while 92% in Case II. If the residual energy chirp is removed, i.e. given a uniform beam energy profile, the final circular polarization degree would increase from 60% to 95% even for Case I. Thus it is convinced that the residual energy chirp, rather than the beam slippage is deleterious to the FEL polarization quality for SDUV-FEL.

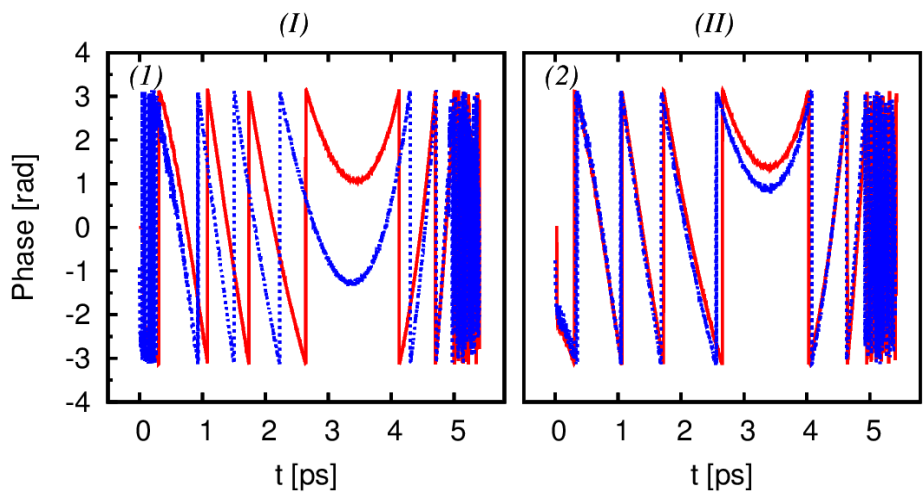

Fig. 4. Radiation phase of $E_x$ (solid line) and $E_y$ (dash line) comparison between Case I (left) and Case II (right).

## 4. The Crossed Undulators and Phase-Shifter

The crossed planar undulator is one of the most important components for FEL polarization control experiments. At present, there is a planar permanent magnetic undulator PMU50-H with alterable gap at SDUV-FEL, which can generate FEL with linear polarization at x-direction, and a new customized PMU50-V whose magnetic field is crossed with the PMU50-H, will generate radiation with linear polarization at y-direction.

The PMU50-V structure is assumed to be similar with the existing PMU50-H except the field direction. The magnetic poles are made of Nd-Fe-B material, and the residual magnetism is greater than 1.23 T. The gap of PMU50-V can be altered by a mechanical-motor from 12 to 60 mm, and the peak field can be up to 0.9 T. The simulation results show that the final polarization performance is not sensitive to the field error of undulator, therefore peak-to-peak field error less than 2% is ensured in the design of PMU50-V.

The phase-shifter is another key component. It is made up of 3 small electromagnetic dipoles, chicane-like structure, with total length of 0.3 m. The dispersion of the phase-shifter is about 2 microns, which enables a phase tuning ability from 0 to $4\pi$ for 523 nm radiation.

## 5. Optical Diagnostics for FEL Polarization

The FEL radiations can be extracted out of the vacuum chamber after the crossed undulators by an optical transition radiation (OTR) target UNPRF3 which is used to observe laser-beam spatial overlap in PMU50-H during the EEHG operation of SDUV-FEL. The reflected angle can be controlled by rotating the OTR mirror, then the pulse energy and spectra can be measured at different angle. At the downstream of the reflection OTR mirror (seen in figure 5), several kinds of optical devices are placed on the optical diagnostic bench, including focusing lens, polarizer, quarter wave-plate and detectors. One can choose different azimuths ($\theta_c$ and $\theta_p$) to obtain the transport matrices and the corresponding light intensities (measured at D2) for getting all Stokes parameters of the FEL radiation mathematically, then figure out the polarization degree and angle according to formulae (1) and (2). Since the accuracy of the polarization measurement highly depends on the intensity stability of the output FEL, a reference pulse energy detector (D1) is utilized for avoiding the shot-to-shot fluctuations induced by the electron beam instability.

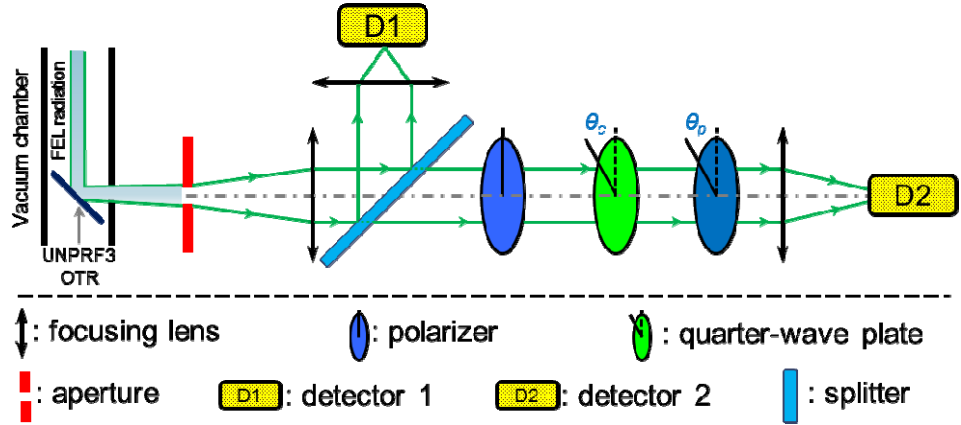

Fig.5. Illustration of the measurement of FEL polarization.

An off-line calibration of the optical diagnostic bench could be accomplished by using a green light, e.g., the second harmonic of the seed laser. Then the on-site measurement can be achieved by tuning the magnetic field of the small phase-shifter between the two radiators, and thus the polarization status can be switched and measured in real time. In the experiment, a movable optical aperture could be utilized to select the region of interest from the whole transverse plane and investigate the spatial non-uniformity effects on the polarization performance.

## 6. Conclusions

With some modifications on the current configuration, FEL polarization control experiment can be carried out at SDUV-FEL. After the electron beam modulated by the 1047 nm laser, a pair of crossed undulators will be used to generate two linear polarized 523 nm radiations, $E_x$ and $E_y$, respectively. Then the polarization status of combined FEL can be controlled with an electromagnetic phase-shifter. At the end of the crossed undulators, optical diagnostic station can provide on-site characterizations, including spectra, pulse energy and the polarization parameters of the output FEL. Start-to-end simulations show that, the total polarization degree can be up to 95% and the circular polarization can be no less than 90% with FEL pulse energy of 100 nJ level.

Moreover, the polarization control experiment at SDUV-FEL serves as a prototype experiment for the Shanghai soft x-ray FEL, in which, high brilliance, short-pulse, fully coherent FEL radiations with variable polarization is expected in the "water-window" spectral regime, which may open up new scientific opportunities in various research fields.

**Acknowledgements**

The authors would like to thank Yuhui Li, Bart Faatz and Zhirong Huang for useful discussions. This work is supported by Natural Science Foundation of China under Grant No. 11175240 and Major State Basic Research Development Program of China under Grant No. 2011CB808300.